\documentclass[]{aa}
\usepackage{longtable,graphicx,amssymb,natbib,rotating,psfrag,amsmath,epsfig,array,txfonts}

\begin{document}

\title{Observations of the intense and ultra-long burst GRB~041219a with the Germanium Spectrometer on
INTEGRAL\thanks{Based on observations with INTEGRAL, an ESA project with instruments and science data centre funded by ESA member states (especially the PI countries: Denmark, France, Germany, Italy, Switzerland, Spain), Czech Republic and Poland, and with the participation of Russia and the USA.} }

\author{
    S.\,McBreen \inst{1}$^,$\inst{2} \and
L.\,Hanlon \inst{3}\and
    S. McGlynn \inst{3} \and
    B.\,McBreen \inst{3} \and
         S. Foley \inst{3} \and
    R.\,Preece\inst{4} \and
    A.\,von\,Kienlin \inst{2} \and
    O.R.\,Williams \inst{5}
}

\offprints{S. McBreen,
\email{smcbreen@mpe.mpg.de}}

\institute{Research and Scientific Support Department of ESA, ESTEC, Noordwijk,
The Netherlands.
 \and  Max-Planck-Institut
  f\"{u}r extraterrestrische Physik, 85748 Garching, Germany.
\and
School of Physics, University College
  Dublin, Dublin 4, Ireland
\and Department of Physics, University of Alabama at
 Huntsville, USA
    \and
  ISOC, ESA/ESAC, Urb. Villafranca del Castillo, Madrid, Spain
  }

\date{Received / Accepted}
\abstract
{GRB~041219a is the brightest burst localised by \textit{INTEGRAL}. The
peak flux  of 
43\,ph\,cm$^{-2}$\,s$^{-1}$ (1.84 $\times 10^{-5}$ ergs\,cm$^{-2}$\,s$^{-1}$, 20\,keV--8\,MeV, 1\,s integration)
is greater than that for $\sim$98\% of all bursts
and the T$_{\rm 90}$ duration of $\sim$186\,s ($\sim$20\,keV--8\,MeV) is longer
than all but a small number of bursts. The intense burst
occurred about $\sim$250~s after the precursor and the long delay
 enabled optical and near infrared telescopes to observe the prompt emission.
We present comprehensive results of the temporal and spectral analyses, including line and afterglow searches  using 
the  spectrometer, SPI, aboard \textit{INTEGRAL},  BAT on \textit{Swift} and
ASM on \textit{Rossi X-ray Timing Explorer}.  
We
avail of multi-wavelength data to generate 
broadband spectra of GRB~041219a and afterglow. 
Spectra for the burst and sub--intervals were fit by the Band model and also by the quasithermal model.
The high resolution
Germanium spectrometer data were searched for emission and absorption features and  for $\gamma$-ray afterglow. 
The overall burst and sub--intervals are well fit by the Band model.
The photon index below the break energy shows a marked change after the
quiescent time interval.  In addition the spectra are well described by a black body component with a power law. 
The burst was detected by BAT and ASM during the 
long quiescent interval in SPI indicating the central
engine might not
be dormant but that the emission
occurs in different bands.
No significant  emission or absorption features were found and limits  of 900~eV and 120~eV are set on the most significant features.
No $\gamma$-ray afterglow was detected from the end of the prompt phase  to $\sim$12~hours post-burst. 
Broadband spectra of the prompt emission were generated in 7 time intervals  using $\gamma$-ray, x-ray, optical and 
near-infrared data and these were compared to the high-redshift burst GRB~050904.
The optical and $\gamma$-ray emission are correlated in GRB~041219a.
We estimate
isotropic radiated energy (E$_{\rm iso}$) to be $\sim$ 5 $\times 10^{52}$erg.
The spectral lag was determined using data from the BAT and it changes throughout the burst.  
A number of pseudo-redshifts were evaluated and large dispersion 
in values was found.
}

\titlerunning{Observations of GRB~041219a with INTEGRAL}
\authorrunning{S. McBreen et al.}

\maketitle
\section{Introduction}
The afterglow era of gamma--ray bursts (GRBs) has yielded many
discoveries, in particular, conclusive proof of the cosmological origin and
association with supernovae \citep[e.g.][]{costa:1997,hjorth2003}.
\textit{INTEGRAL} has detected 37 GRBs to date  
most of which were quite weak  including a member of the low-luminosity class of bursts GRB~031203 \citep{sls04,watson2004} and an x-ray rich burst GRB~040223 \citep{sinead2005}.
For recent reviews of $\gamma$-ray bursts
see \citet{Zhang2004} and \citet{piran:1143}.
In this paper we present the results of the $\gamma$--ray spectral and
temporal characteristics of the burst GRB~041219a obtained with the high
resolution spectrometer, SPI,  aboard \textit{INTEGRAL} \citep{wink2003}.
SPI is a coded-mask telescope with a fully coded field of view of 16$\degr$
that uses a 19 pixel high spectral resolution Ge detector. The detectors cover
the energy range 20~keV--8~MeV with an energy resolution of 2-8~keV FWHM. A detailed description of SPI can be found in \citet{ved2003}.
 Data from the imager IBIS is not included here because proprietary rights were awarded to other groups.
Analysis of the light curves observed by Burst Alert Telescope (BAT) on \textit{Swift}
 in the range 15 --350\,keV  \citep{gehrels:2004} and the
 {Rossi X-ray Timing Explorer} All Sky Monitor (ASM)  in the energy range 1.5--12~keV  \citep{LEvine1996} 
  are also presented.
  GRB~041219a is the brightest burst
detected by the \textit{INTEGRAL}  burst alert system (IBAS) \citep{mgbwp03} and prompt emission was  detected in the optical \citep{vestrand:2005} and near infrared \citep{blake:2005}.
\psfrag{UNIT}[c]{keV (  keV cm$^{-2}$ s$^{-1}$ keV$^{-1}$)}
   \begin{figure*}[t]
  \begin{center}
  \includegraphics[width=0.99\textwidth]{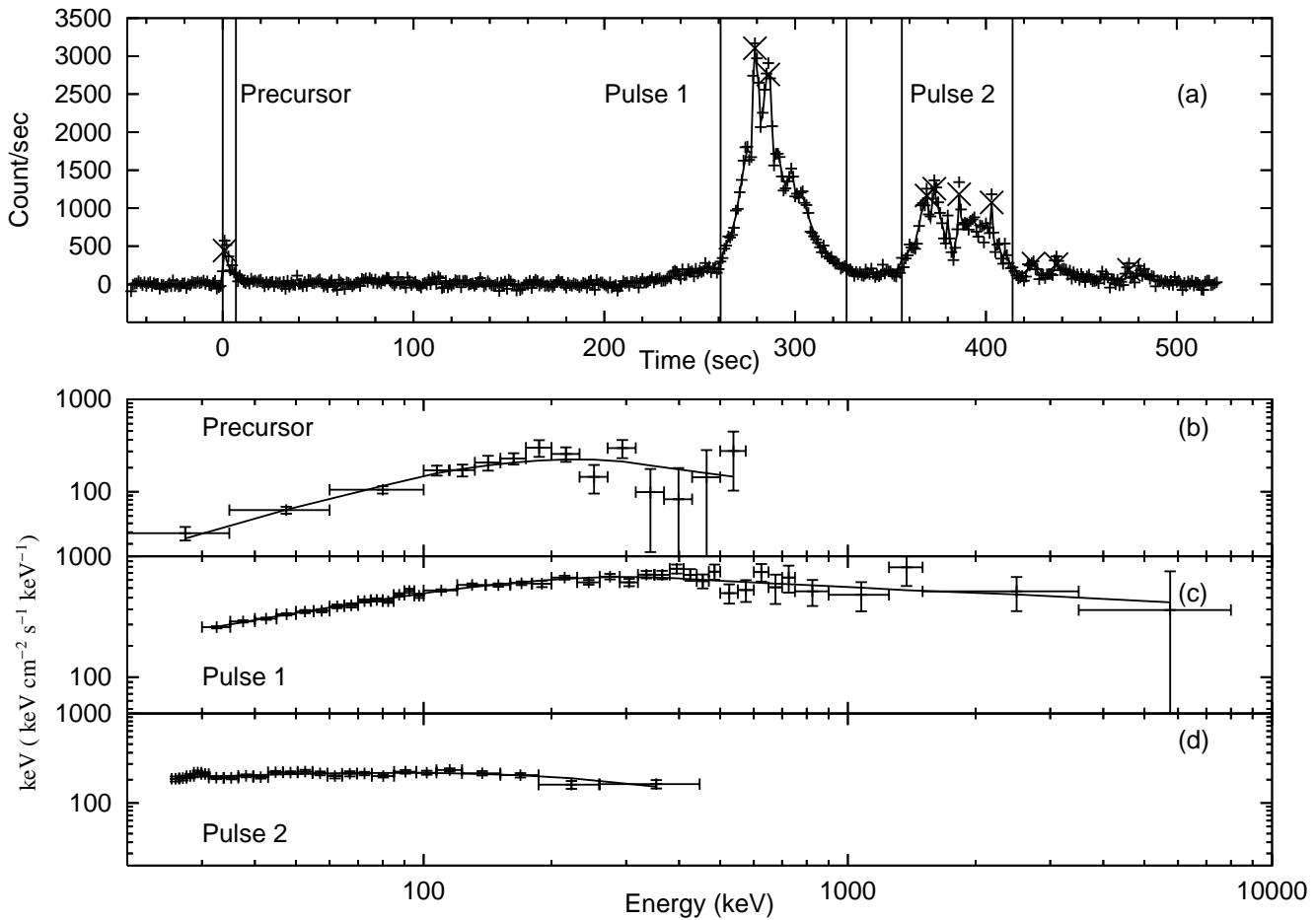}
    \caption{SPI light curves and spectra of GRB\,041219a.
    (a) denoised and background subtracted light curve where significant
pulses  selected  by the algorithm are marked with an \textit{x}.
The three marked sections denote the time intervals over which
the spectra were analysed. The start and end times of the intervals
and the parameters of
best fit Band spectra are presented in Table 1.
 Spectrum and Band model fits for 
(b) the precursor pulse (0--7~s);
(c)  pulse 1 (261--327~s);
(d) pulse 2 (356--414~s).
\label{spilc}
}
  \end{center}
\end{figure*}

\section{Observations}
GRB\,041219a was detected by IBAS at 01:42:18\,UTC on
December\,19$^{\rm th}$ 2004 \citep{gms+2004} at a location of
right ascension 00h 24m 25.8s, declination +62$^{\degr}$ 50${^\prime}$
 05.6$^{\prime\prime}$
  (Galactic latitude and longitude of
0.12$\degr$, 119.85$\degr$)
at a detector off-axis angle of 3.2$\degr$.
GRB~041219a was also detected by BAT  \citep{bbc+2004}
at a location consistent with the IBAS position.
\citet{ASMGCN} reported a serendipitous observation of GRB~041219a during two 90~s dwell times with the \textit{Rossi X-ray Timing Explorer} All Sky Monitor (ASM) in the energy range 1.5--12~keV  \citep{LEvine1996}. The burst was in the field of view from 
6~s after the trigger for 3 minutes apart from a 6~s gap. 
Prompt optical emission correlated with the $\gamma$-ray emission was reported by \citet{vestrand:2005}.
Infrared emission was detected 7.2 minutes after the burst trigger
at the tail end of the prompt $\gamma$-ray emission \citep{blake:2005}.
Radio observations detected a source at 8.5~GHz at T$_{\rm 0} +$ 1.1~days
 \citep{Soderberg04}
and two observations at 4.9~GHz beginning at T$_{\rm 0}+$1.59~days 
\citep{van1:04}
and T$_{\rm 0}+$2.54 by \citep{van2:04} reported
an increased flux density in the second epoch.
There is no measured redshift available for this burst.

\begin{table*}[width=0.6\columnwidth] %[t] %[B]
\caption{Spectral properties of  GRB\,041219a. 
The columns refer
to the emission region, time interval of spectral fits,  low-energy power-law index ($\alpha$), high-energy power-law index ($\beta$),  break energy (E$_{0}$),
\textit {$\chi^2$}/degrees of freedom (dof)
 of the fit, fluence in the range 20--200~keV (S$_{\rm 20-200~keV}$), fluence in the range 20~keV--8~MeV (S$_{\rm 20~keV-8~MeV}$),
and value of the spectral lag between BAT channels 1 and 3 over the specified interval.  Errors on spectral parameters quoted for a 90\% confidence level and the fit parameters are for the spectra shown in Fig.~\ref{spilc}.
\label{spec}
}
\centering
\begin{tabular}{c|c|c|c|c|c|c|c|c}
\hline\hline
Emission &  Time~(s) &$\alpha$   & $\beta$ &
E$_{0}$  &  \it $\chi^2$/dof   & S$_{\rm 20-200~keV}$
 & S$_{\rm 20~keV-8~MeV}$ & Lag~(s) \\
Region &  &   &  & (keV) &  & (erg cm$^{-2}$ ) &  (erg cm$^{-2}$ )& \\
\hline
 Precursor&0--7 & $-$0.45 $^{+0.37}_{-0.30}$ & $-$2.62 $^{+0.73}_{-7.4}$ & 145.4 $^{+79.7}_{-48.1}$ &  12.1/11 &  2.8$\times 10^{-6}$ &
 7.1 $ \times 10^{-6}$  &  1.2$\pm$0.08\\%/0--7\\
 1$^{\rm st}$ Pulse &261--327 & $-$1.5$^{+0.08}_{-0.06}$ &  $-$1.95 $^{+0.08}_{-0.21}$ & 568.3 $^{+310}_{-205.2}$ & 57.5/50 &  7.0$\times 10^{-5}$  &  2.6$\times 10^{-4}$ & 0.11s$\pm$0.01\\%/275--290 \\
 2$^{\rm nd}$ Pulse& 356--414 & $-$1.76 $^{+0.09}_{-0.08}$ & $-$3$^\dagger$ & %$^{+0.15}_{-0.38}$ &
363.6 $^{+193.1}_{-99.5}$ & 30.9/21 & 4.0$\times 10^{-5}$&
1.0$ \times 10^{-4}$  & 0.15$\pm$0.01\\%/350--400 \\
 1$^{\rm st}$ \&  2$^{\rm nd}$ Pulse & 261--414& $-$1.43 $^{+0.08}_{-0.06}$ & $-$2.06 $^{+0.09}_{-0.12}$ &
299.2 $^{+76.2}_{-73.9}$ & 16.5/25  &  1.3$\times 10^{-4}$ &
3.7 $\times 10^{-4}$ & 0.17$\pm$0.02\\%/250--450 \\
 Burst &0--580 & $-$1.48 $^{+0.08}_{-0.07}$ & $-$1.92 $^{+0.07}_{-0.13}$ & 365.9 $^{+191.6}_{-108.2}$ &  51.9/30 &  1.6$\times 10^{-4}$ &
 5.7$ \times 10^{-4}$ & \\
\hline
\end{tabular}
 \begin{minipage}[c]{0.99\columnwidth}
 \vspace{2em}
 \end{minipage}
  \begin{minipage}[c]{0.99\columnwidth} 
   {\hspace{-12em} \footnotesize{$^\dagger$  $\beta$ value frozen at $-$3.}}
 \end{minipage}
\end{table*}

 \section{Analysis and Results}

\subsection{Light Curves}
\begin{figure}[htbp]
\begin{center}
\includegraphics[width=\columnwidth]{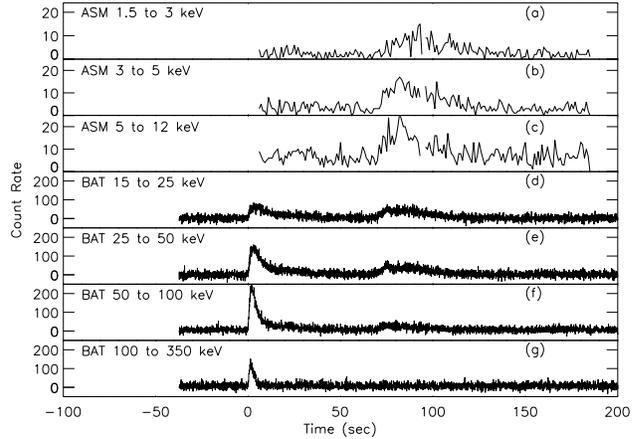}
\caption{ \textit{Rossi} ASM (a-c) and \textit{Swift} BAT (d-g) light curves of GRB~041219a up to T$_{\rm 0}+200$~s. \textit{Rossi} ASM data are available only for two 90~s 
dwell times starting at T$_{\rm 0}+6$~s and T$_{\rm 0}+96$~s.
The ASM and BAT light curves have emission in the interval at about 80~s
that is  quiescent in the SPI light curve (
Fig.~\protect{\ref{spilc}}). The \textit{Rossi}-ASM data were obtained
courtesy of  Alan Levine.
\label{BATcurve}}
\end{center}
\end{figure}

The SPI light curve  in the broad
energy band from 20\,keV\,--\,8\,MeV at 1~s resolution
is given in Fig.~\ref{spilc}~(a).  The light curves are generated from housekeeping data \citep{moran2004}.
A section of the light curve detected by the BAT is shown in Fig.~\ref{BATcurve} using the data from \citet{fbc+2004} and no spectral information is available.
The T$_{\rm 90}$ value of 186\,s  ($\sim$20~keV--8~MeV) 
was determined  from the SPI  light curve.
T$_{\rm  90}$ is the time for 5 to 95\% of the flux to be accumulated and
the values are shorter than the burst duration in this case because of the
quiescent interval and the precursor has only ~ $\sim$2\% of the counts.

The temporal structure of the burst is 
unusual, with an initial weak precursor pulse
followed  by a long quiescent time interval
and the main emission beginning at $\sim$ 250~s post--trigger (Fig.~\ref{spilc}~(a)).
The SPI light curve was denoised using a wavelet analysis
\citep{quilligan:2002} and the pulses selected by a pulse decomposition algorithm are shown in Fig.~\ref{spilc}~(a).

Although the SPI light curve is quiescent from $\sim$7--200~s, emission is detected in the BAT light curve
 particularly in the lower energy channels 1 and 2 (Fig.\,\ref{BATcurve}) (15-25~keV  and 25- 50~keV) \citep{fbc+2004}.  In addition a spectrally soft pulse was detected in the ASM \citep{ASMGCN} 
beginning at 80~s  in the quiescent period of the SPI observation
(Fig.~\ref{BATcurve}). 
{The value of T$_{\rm 90}$ depends very much on the 
bandwidth over which it is measured.}

The spectral lags and associated
errors were measured between BAT channels 1 and 3  using the
cross-correlation technique described in \citet{nmb00}.
The code was validated by comparing results for a test case GRB
obtained by Norris et al. and consistency within the errors was obtained,
over a range of parameters.
The values for the lag in Table~\ref{spec} (Column 9)
 were measured for regions where  the intensity was greater
than 0.1$\times$ peak intensity of all channels combined. The timing data were
oversampled by a factor of 10 for the main emission and a factor of 4 for
the precursor in order to allow bootstrap errors to be calculated.
The maximum of the cubic fit to the cross-correlation function was used as the
measure of the spectral lag.   The lag in the precursor  is much longer than
 that in the remainder of the burst (Table\,\ref{spec}).

\subsection{Spectral Analysis}

The spectrum of GRB\,041219a was extracted using
Online Software Analysis version 5.0 
available from the \textit{INTEGRAL Data Science Centre} \footnote{http://isdc.unige.ch/} (see also Diehl et al 2003).%\citet[][see also]{Diehl2003}.  
GRB~041219a is the brightest burst localised by \textit{INTEGRAL} to date. Its
peak flux  of
43\,ph\,cm$^{-2}$\,s$^{-1}$ (1.84$\times 10^{-5}$ ergs\,cm$^{-2}$\,s$^{-1}$)
(20\,keV--8\,MeV, 1\,s integration)  is greater than $\sim$98\% of bursts detected
by BATSE while its T$_{\rm 90}$ duration of 186\,s ($\sim$20\,keV--8\,MeV) is longer
than all but a handful ($\sim$4\%) of BATSE GRBs \citep{Paciesas:1999}.
The spectrum of the burst and sub-intervals are well fit by the Band model \citep{band:1993} and the parameters are presented in Fig.~1 and Table~\ref{spec}~(Columns 1--8).
The parameters of the spectrum evolve during the burst.
For instance  $\alpha$ is remarkably higher in the initial pulse in comparison to  the main emission phase.
The  value of $\alpha$ for the main emission phase is
$-1.43^{+0.08}_{-0.06}$ and is used to derive the redshift in
$\S$~\ref{x_section}.
The peak energy, E$_{\rm peak}$, is given by
$(2+\alpha) \times \,E_{\rm 0} $ and
the evolution of E$_{\rm peak}$Ê
shows softening in the main emission phase.
The value of E$_{\rm peak}$ in the interval of the main emission phase is 170$^{+44.5}_{-42.6}$~keV and this value is used to derive the radiated and isotropic energy  in $\S$~\ref{x_section}.

\begin{figure}[h]
\begin{center}
\psfrag{UNIT}[c]{keV (  keV cm$^{-2}$ s$^{-1}$ keV$^{-1}$)}
\includegraphics[width=\columnwidth]{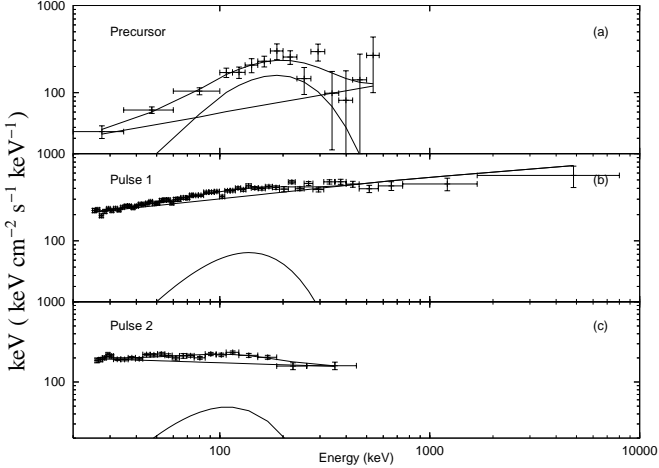}
\caption{Spectral fits to (a) the precursor, (b) pulse 1 and
(c) pulse 2 of GRB\,041219a using combined black body and power law fits. The data are shown along with the combined fit
and the black body and power law components.
\label{bbpo_fig}}
\end{center}
\end{figure}

\citet{ryde05b} suggested that thermal emission may be ubiquitous in GRBs and fit a black body + power law model to a number of bright BATSE
bursts. \citet{ryde05b} reported that the temperature was intially constant or weakly decreasing with a shallow power law and that the power law index steepened throughout the burst. We fit the quasithermal model to GRB\,041219a and the results are shown in Table\,\ref{bbpo_spec} and Fig.\,\ref{bbpo_fig}.  In this interpretation the black body contributes more strongly in the precursor than in pulses 1 and 2.

 \begin{table*}[ht]%[hb] %[B]
\caption{
Spectral fits to GRB\,041219a using the quasithermal model.
The columns refer
to the emission region, time interval of spectral fits,  \textit{kT},
 power-law index ($\Gamma$),
\textit {$\chi^2$}/degrees of freedom (dof)
 of the fit.
  Errors on spectral parameters quoted for a 90\% confidence level.
\label{bbpo_spec}
}
\centering
\begin{tabular}{c|c|c|c|c|c|c}
\hline\hline
Emission &  Time~(s) &  \textit{kT} & $\Gamma$ &  \it $\chi^2$/dof  & 
$S_{20-200~keV}$ in the blackbody & \% Flux in the blackbody \\
Region &                &   & &   & component (ergs cm$^{-2}$ s$^{-1}$)& component (20-200~keV)\\
\hline
 Precursor&0--7 &  45.7$^{+9.1}_{-8.7}$ & 1.58$^{+0.53}_{-0.28}$ & 9.25/11
 & 1.4$\times 10^{-6}$ & 49\%\\
 1$^{\rm st}$ Pulse &261--327 &  35.1$^{+5.5}_{-4.7}$ & 1.74$^{+0.02}_{-0.02}$ & 68/50 & 8.7 $\times 10^{-6}$& 11.6\%\\
 2$^{\rm nd}$ Pulse& 356--414 & 27.3$^{+4.9}_{-4.8}$& 2.1$^{+0.70}_{-0.08}$ & 28.9/21 & 5.8$\times 10^{-6}$ & 12.7\%\\
\hline

\end{tabular}
\end{table*}

\subsection{Spectral lines}

A search for line emission in the SPI data was carried out on the brightest pulse in the
burst.  The search  involved
finding
 the best continuum model for the data and then adding a gaussian emission
 line of varying width (1~keV to 20~keV)
to the model at energies  from 30~keV to 1~MeV.
 The F-test was used to evaluate the resulting improvement in the fit 
 for each line width and each energy.
 However, it is well known that the F-test alone cannot be used to check for
the presence of a line \citep{prot2002}
and must be calibrated for false positives.
We adopt a similar approach to that
suggested by  \citet{prot2002}. % ($\S$~5).
The spectrum for which the largest F-test value was obtained was chosen and
Monte Carlo simulations were carried out to test
the number of times a more significant F-test value
would be detected in simulated data of the continuum model.
A large number of spectra (10,000) of the continuum were generated.
The Band model and Band model+ emission line
were fit to each generated spectrum and the F-test for the improvement
in the fit was recorded for each.
 The same approach was adopted to search for absorption
features. 
The false positive rate was evaluated.

The most significant emission line feature was 4~keV wide  at 89~keV
with an F-test value of 4.6.
Of the 10,000 simulated spectra, 5850
resulted in an improvement in $\chi$$^2$
and out of this fraction
 5.2\% of trials resulted in a higher value of F-test than 4.6.
The  equivalent width of this emission line
is  900~eV (1.4$\times 10^{-9}$ ergs).

The most significant absorption feature had a width of  3~keV at
103~keV with an F-test value of 7.8. Of the 10,000
simulations that were run, 5876 resulted in an improvement
in $\chi^2$ and 2.5\% of these produced an F-test value
higher than 7.8.
The equivalent width of this absorption feature
is 120~eV (1.9$\times 10^{-10}$ ergs).

No significant emission or absorption features were found. 

\subsection{Afterglow Search with SPI}
\begin{figure}[h]
\begin{center}
{\includegraphics[width=\columnwidth]{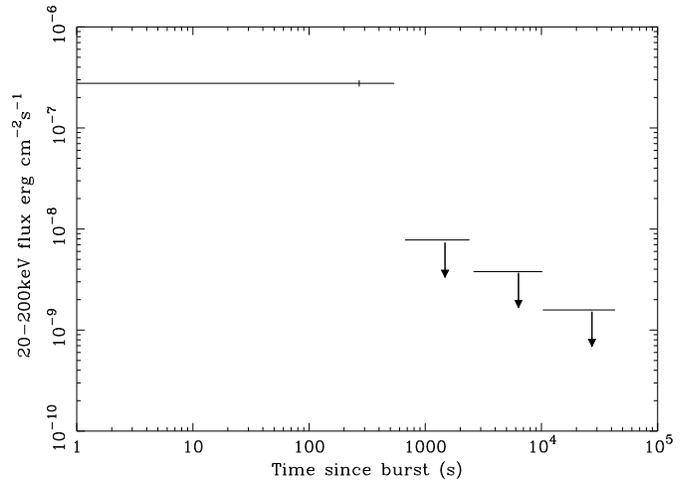}}
\caption{The flux averaged over the burst duration of GRB~041219a
and  3 $\sigma$ upper limits in the energy range 20--200~keV.
The arrows give the 3 $\sigma$ limits over the time 670~s to
43000~s after the burst.
\label{limitss}}
\end{center}
\end{figure}

A search for a $\gamma$--ray afterglow was performed on the available data for
up to 12 hours after the burst trigger. The data was split into three
separate time intervals and a spectrum extracted.
Significant emission from the  burst position was not detected even when the data were summed over the entire 12 hours. 
The 3 $\sigma$ upper limits for the subsequent time
intervals are given in Fig.\,\ref{limitss}.

\subsection{Broadband spectra} %of prompt emission}

\begin{figure}[htbp] %[htbp]
\begin{center}
\psfrag{F}[c]{F$_\nu$ (ergs cm$^{-2}$s$^{-1}$Hz$^{-1}$)}
\psfrag{Energy}{Energy (keV)}
\psfrag{A:}{\tiny A: T$_{\rm 0}$ to T$_{\rm 0}$+7s}
\psfrag{B:}{\tiny B: T$_{\rm 0}$+8~s to T$_{\rm 0}$+65~s}
\psfrag{C:}{\tiny C: T$_{\rm 0}$+66~s to T$_{\rm 0}$+120~s}
{\includegraphics[width=\columnwidth]{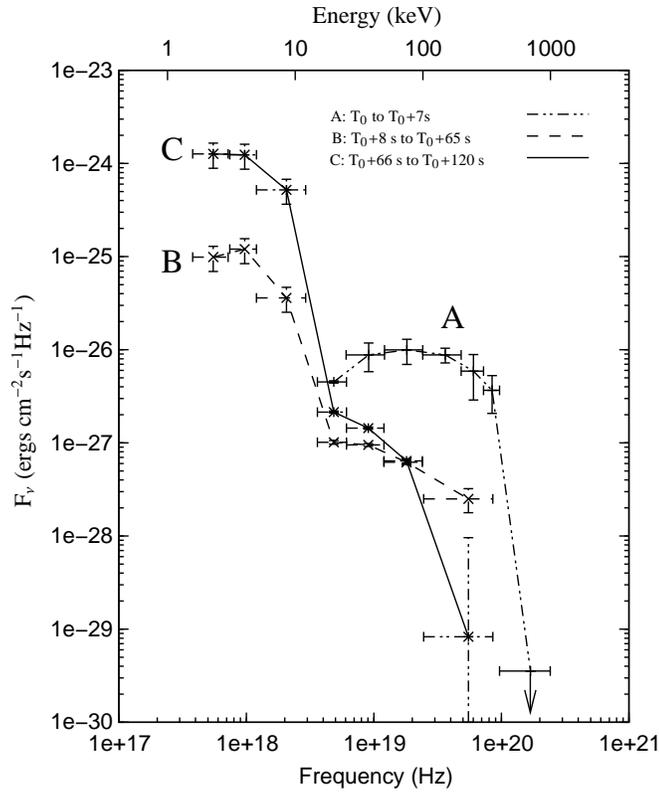}}
\caption{
Broadband spectra of the initial 120~s of GRB~041219a in the x-ray and gamma ray region.  The data are plotted in flux density (ergs cm$^{-1}$ s$^{-1}$Hz$^{-1}$) versus observed frequency (lower axis) and keV (upper axis).   The first interval (T$_{\rm 0}$ to  T$_{\rm 0}+$7~s) includes the precursor pulse and the third interval T$_{\rm 0}+$66~s to  T$_{\rm 0}+$120~s includes the soft pulse in the long quiescent interval.  The broad spectra were generated using ASM x-ray data and $\gamma$-ray data from BAT and SPI.
\label{asmSED}}
\end{center}
\end{figure}

The broadband spectra of the prompt emission for the GRB~041219a
are plotted in 7 time intervals using a combination of $\gamma$-ray data 
from SPI and BAT, x-ray data from ASM, optical data from 
\citet{vestrand:2005} and infrared data from 
\citet{blake:2005}  {corrected for extinction}.
Fig.~\ref{asmSED} shows three spectra during the initial $\sim$120~s generated 
from \textit{Rossi}-ASM, \textit{Swift}-BAT and \textit{INTEGRAL}-SPI data in three different time
intervals.

\begin{figure}[htbp]
\begin{center}
\psfrag{Hz}{Frequency (Hz)}
\psfrag{y}[c]{F$_\nu$ (ergs cm$^{-2}$s${^-1}$Hz$^{-1}$)}
\psfrag{A:}{\tiny A: T$_{\rm 0}$+203~s to T$_{\rm 0}$+275~s }
\psfrag{B:}{\tiny B: T$_{\rm 0}$+288~s to T$_{\rm 0}$ +318~s}
\psfrag{C:}{\tiny C: T$_{\rm 0}$+330~s to T$_{\rm 0}$ +403~s}
\psfrag{D:}{\tiny D: T$_{\rm 0}$+415~s to T$_{\rm 0}$ +573~s}
\psfrag{F:}{\tiny F:T$_{\rm 0}$+151~s to T$_{\rm 0}$ +254~s}
\psfrag{E:}{\tiny E:T$_{\rm 0}$+12~hours}
{\includegraphics[width=\columnwidth]{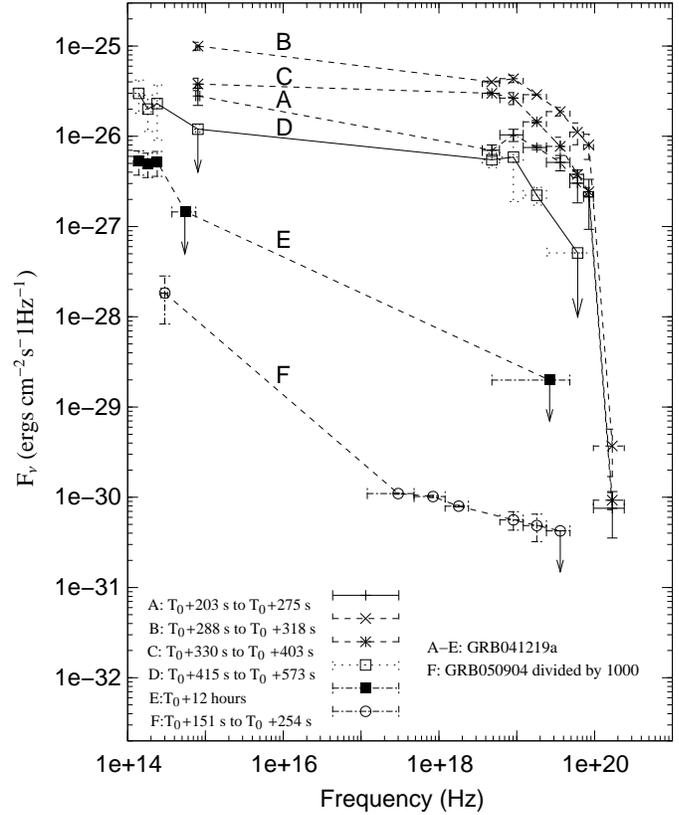}}

\caption{
Broadband spectra of GRB~041219a (A to E) and GRB~050904 (F) at a redshift of 6.29.  The data are plotted in flux density (ergs cm$^{-2}$ s$^{-1}$ Hz$^{-1}$) versus observed frequency.  The four intervals (A to D) are defined by the periods of simultaneous RAPTOR optical observations of GRB~041219a (Vestrand et al. 2005).  The RAPTOR optical results, corrected for extinction, are shown and connected by a line to the lowest energy BAT channel (15 - 25 keV) and the higher energy SPI channels.  The PARITEL near infrared results (Blake et al. 2005) are shown for interval D and connect by a solid line with the optical upper limit and the $\gamma$-ray data.  The broadband spectrum of the afterglow from GRB~041219a is shown 12~hours after the burst (E) using infrared and optical data from \protect\citet{blake:2005} and 
 \protect\citet{2004Sonda} respectively and
 and an upper limit from the afterglow search with SPI. 
  The broadband spectra show the evolution from the peak of the burst in B through the subsequent intervals (C and D) and into the afterglow.  The simultaneous optical, x-ray \protect\citep{boer2006} and $\gamma$-ray results (this work)
 for the high redshift GRB~050904 
are shown for the interval 151 to 254~s.  The data for GRB~050904 have been divided by 10$^3$ for clarity of presentation and 
unlike GRB~041219a, GRB~050904 clearly shows the large difference between
the optical and $\gamma$-ray emission.
 \label{SED}}
\end{center}
\end{figure}

The broadband spectra of the prompt emission for GRB~041219a are given in Fig.~\ref{SED} for the four time intervals of simultaneous optical and gamma
observations.  In the case of the interval (D) there are also simultaneous near infrared results from PARITEL \citep{blake:2005}.  The broadband spectra of GRB~041219a were originally presented by Vestrand et al. (2005).  The energy range has been expanded here by including the SPI results for intervals (A) to (D) and near infrared results for interval (D). 
The broadband spectrum of the afterglow from GRB~041219a is shown 12~hours after the burst (E) using infrared and optical data from \citet{blake:2005} and 
\citet{2004Sonda} respectively  and an upper limit from SPI.  The simultaneous optical, x-ray \citep{boer2006} and gamma ray results %(this work)
 for the high redshift GRB~050904 
are shown in the interval 151 to 254~s for comparison.

\section{Discussion}
\subsection {Spectral behaviour of GRB~041219a}
In most GRB models, the main event is preceded by  less intense emission, characterised by a thermal spectrum, called a precursor. Precursors are associated with  the transition of the fireball from an optically thick
to an optically thin environment  \citep[e.g.][]{piran:1143,rr_macf_lazz:2002}.
A rigorous definition of precursors is
 problematic because of the complexity of GRB temporal structures.
 \citet{koshut:1995} defined precursors that had a peak intensity
 lower than that of the whole burst and were followed by a period of quiescence longer than the remaining duration of the burst.
Soft precursors, occurring before the trigger, have been detected by
many instruments e.g.
\textit{Ginga} \citep{murk:1991}, \textit{HETE} \citep{2004Van},
and BATSE \citep[e.g.][]{2005Laz}.
The \textit{Ginga} precursor has a thermal spectrum and has
been compared  with theoretical expectations \citep{rr_macf_lazz:2002}.
The precursor pulse was fit by a Band model and a black blody+power law
model  and is quite different from the main emission. The properties of the precursor spectrum are shown in Tables~\ref{spec} and \ref{bbpo_spec} respectively.

The $\beta$ values for a simple cooling spectrum and an instantaneous
spectrum, in which cooling effects are negligible, are
$-\frac{(p+2)}{2}$
and $-\frac{(p+1)}{2}$ respectively, where $p$ is the electron index.
From Table 1, using the well constrained $\beta$ values for the
main emission phase, the derived $p$ values are $2.12^{+0.24}_{-0.18}$ and
$3.12^{+0.18}_{-0.24}$.

The value of  $\alpha$ changes significantlyÊ between the precursor
and
main emission phase. $\alpha$ is predicted by the synchrotron shock
model to be between $-3/2$ and $-1/2$, although taking into account
non-isotropic electron pitch angles and/or self-absorption can produce
$\alpha$ values up to and above 0 \citep{lloyd2000}.
The evolution of $\alpha$ may be evidence of changes in opacity or
electron pitch angle from pulse to pulse. There is a tendency for low
break
energies to give harder $\alpha$ values, since the spectrum does not
reach
the asymptotic value in the limited spectral fitting window. 
 
 The majority of GRBs have low energy power law spectral indices in the range 
 $-3/2 < \alpha < -2/3 $ \citep[e.g.][]{crider1997,pbm+00,ghirlanda2002}. However a substantial fraction of GRBs
  have $\alpha >  -2/3$ and lie outside the 
 optically thin synchrotron model \citep{preece_sync}.  A number of explanations have been 
 offered such as jitter radiation \citep{med2000},  synchrotron emission from 
 particles with an anisotropic pitch angle \citep{lloyd2000,LL_RR:2002}, 
 synchrotron self-Compton or inverse Compton off photospheric photons
 \citep{mes2000},  Compton drag \citep[e.g.][]{2000lazzati}, Comptonisation 
 of low energy photons \citep[e.g.][]{1997liang}.
 
 A number of authors have discussed the possibility of fitting thermal components to time resolved or average GRB spectra \citep[e.g.][]{ghirlanda2003, Kaneko2003,ryde2004,ryde05b,bosnjak2006}. %,ryde_conf}.
 \citet{ghirlanda2003} studied time resolved spectra of a number of BATSE 
 bursts with very hard spectra and found that the spectra 
 were not adequately explained by non-thermal emission models. In fact, 
 they found that  spectra early in the burst were well fit by
 a black body spectral model. 
 We find an improvement in $\chi^2$ for the blackbody+power law
 model over the Band model for the precursor phase of GRB~041219a 
 where 49\% of the flux is contained in the blackbody component
 (Table~\ref{bbpo_spec}).
 The Band model provides a better description of the
 continuum spectrum in pulse 1 and similar goodness of fit values 
 are achieved for pulse 2. 
 The fits of the quasithermal model show that 
 the value of \textit{kT} is highest for the precursor pulse and declines
 during the burst. 
 \citet{Kaneko2003} reported
 BATSE-BeppoSAX joint analysis of GRB~970111
 suggesting existence of a blackbody component with temperature 
 $\sim$40-70~keV in the first  5 seconds of this burst in agreement
 with GRB~041219a.
 Recently,  \citet{ryde05b} studied the
 prompt emission from 25 of the brightest pulses in the 
 catalogue of \citet{dan2003} and found that time resolved spectra
 were equally well  fitted by the black body+power law model 
 and with the Band model. In addition,  \citet{ryde05b} found that
  the thermal and power law components evolve together with 
  the black body temperature decreasing with time. The spectrum 
 of GRB~041219a  can also be described by the quasithermal model
(Table~\ref{bbpo_spec}).

 Recently, \citet{rees2005} suggested that the E$_{\rm peak}$ in the $\gamma$-ray  spectrum is due to the Comptonized thermal component from the photosphere,
 where the comoving optical depth falls to unity.
 The thermal emission from a laminar and steady jet when viewed head-on,
 would give rise to a thermal spectrum peaking in the
 x-ray or $\gamma$-ray band.  The resulting spectrum would be
 the superposition of the Comptonized thermal component and the 
 power law from the synchrotron emission.

 \subsection{Spectral lines}

 The issue of narrow spectral lines in the prompt emission from
GRBs has been controversial. The line features originally
identified in spectra at $\sim 20-60$\,keV \citep{maz80} were
interpreted as cyclotron line emission from electrons in neutron
star magnetic fields and were used to support the neutron star origin
hypothesis for GRBs. The BATSE Spectroscopy Detectors were
 determined to be quite capable
of detecting features \citep{bandetal95}, such as had been reported
by {\em Ginga} \citep{murakamietal88}, if they were a common occurance. The
results from an automated line search of BATSE data, however, were
inconclusive \citep{briggsetal97}.
Our analysis revealed no significant emission or absorption features
in GRB~041219a.

In the context of what is now understood about the progenitors and
distances of GRBs, the explanation for narrow $\gamma$-ray
spectral features in the prompt emission is necessarily different
to the original hypothesis and may be much more complex.
Nevertheless, spectral line features are an extremely powerful
probe of the burst environment. This has been amply illustrated by
the analyses and interpretations of transient X--ray Fe features
and absorption edges in both the prompt emission and afterglows of
some GRBs (see e.g. \citet{bott04} for a review) which have been
used to impose general physics constraints on the homogeneity,
isotropy and location of the reprocessing material with respect to
the burst source. No strong evidence for $\gamma$-ray 
emission or absorption features has been found in the brightest
pulse of GRB~041219a.

 \subsection{Temporal behaviour GRB~041219a}
 GRB~041219a has an unusual time history because the initial pulse
is followed by a long period of quiescence. 
However it should be
noted that although the time interval is quiescent  in the SPI light curve, emission is detected in the three \textit{Rossi}-ASM and 
\textit{Swift}-BAT light curves (Fig.~\ref{BATcurve}) at $\sim$T$_{\rm 0}+$80~s.
Similar light curves with a weak triggering pulse followed by a long quiescent time interval have been observed by BATSE \citep{fish1994}. % (REF).
The sample of \citet{quilligan:2002} contained 319 bright BATSE GRBs and two of the bursts show similar temporal evolution to GRB~041219a (triggers 6451 and 7575). 
In addition the $\gamma$-ray light curves of GRB~050820a 
\citep{2005Page}
and GRB~060124 
\citep{Holland060124, ramano2005} 
show a similar weak precursor pulse 
and a long quiescent interval before the main pulse as noted by \citet{golenetskii2005} and \citet{Golenstskii2006}.

\citet{2005Laz} searched for 'non-triggering' precursors in
a sample of bright, long BATSE  burst  light curves and found
that in  20\% of cases  there is evidence of emission above the background coming from the same direction as the GRB.
This emission is characterized by a softer spectrum with respect to the burst spectrum and contains up  to 1\% of the counts
and typically has a non-thermal power-law spectrum.
The precursor type pulse in GRB~041219a contains
2\% of the total fluence (50--300~keV) and is similar to
precursors in the BATSE sample, except that it was a triggering event.

A number of studies have been made of periods of quiescence in GRBs. A trend was found between the duration of the quiescent time and duration of the following emission period
\citep{RR_1:2000}  that is consistent with GRB~041219a.Ê
\citet{quilligan:2002} showed that
the measured distribution of time intervals between
pulses in BATSE bursts is  best fit by lognormal with allowance made for the excess in time intervals $>$ 15~s.
The Pareto L\'{e}vy tail  { \citep{montshles:1982}  of the time
intervals is well fit by a power-law of slope $\sim-$1 \citep{quilligan:2002}.}

 The model of a GRB as a relaxation system
\citep{mmhq2002}, which continuously accumulates energy and discontinuously releases it accounts for the correlated pulses properties and time intervals between pulses
\citep{mmqh02,nakar:2002}. This model can be extended to include periods of quiescence if the system returns to a more stable configuration that might be caused by a total release of accumulated energy.Ê These general considerations do not identify the emission mechanism.Ê Periods of quiescence can in principle, be caused by a modulated relativistic wind or a switching off of the central engine
\citep{RR_2:2000}. 

\citet{2005Drago}  found that similarities between the pre- and post-emission periods suggest that both emission periods are produced by the same mechanism and that long quiescent intervals are generated by a switching off of the engine rather than a modulation of a continuous wind.

{However, it should be noted that
GRB~041219a was detected at lower energies
 by BAT and ASM during the 
long quiescent interval in SPI indicating the central
engine might not
be dormant but that the emission
occurs in different bands.}

\subsection{Spectral lags}
The results on the spectral lags in GRB~041219a are presented in 
Table~\ref{spec}.
{ The spectral lags of GRBs and their evolution are vital  tools for
probing the emission mechanism, however their interpretation is not
straightforward. 

The empirical connection between the rate of spectral evolution
and spectral lag should ultimately reveal the underlying physical
mechanism for the lag-luminosity relation \citep{kl03}. The lag
correlates linearly with the decay timescale of single pulse
bursts and hence with the peak-energy decay \citep{ryde05}. This
implies that the lag-luminosity relation should translate
into one involving the pulse timescale and hence to the processes
forming the pulse in the outflowing plasma. The distillation of
the spectral lag into a combination of spectral and temporal pulse
properties leads naturally to the conclusion that lag should
evolve during the burst. 
{\citet{hakkila2004}
reported BATSE bursts with greater spectral lags after a long
quiescent interval. \citet{Ryde_Kov:2005} also reported lags differing considerably within bursts.% \citet{2006Zhang} }
These changing lags have been observed in a sample of
BATSE bursts \citep{clw+05} but with no obvious variation in lag as
a function of individual pulse luminosity.} In the case of GRB\,041219a, there is a relatively long lag in the precursor type pulse (Table~\ref{spec}), with much shorter lags determined
in the main emission phase.
The pulses are slower in the first half of the burst, split by the tallest pulse in the burst, using the BAT four channel data. 
{The fact the lag is not constant throughout the burst
has obvious consequences for its use a luminosity indictor.}
%The average pulse width in the first/second half of the burst
%is 11~s/4~s. However the tallest pulse in the burst is the fourth pulse so there are limited statistics in the first half.

\subsection{Constraints on redshift and energy}
\label{x_section}

\citet{aft+02} derived a relationship between $\alpha$ and redshift
from {\it BeppoSAX} $\gamma$--ray
 bursts with known redshifts
which reflects a dependence of $\alpha$ on the E$_{\rm peak}$.
Using the value of $\alpha=$ $-1.43^{+0.08}_{-0.06}$
 from all sub--intervals
in the main emission phase  yields z$=$ 1.43$^{+0.55}_{-0.38}$. % (Eq.\ref{eq:amati}).
The relationship between E$_{\rm peak}$ and E$_{\rm iso}$ \citep{aft+02,ghirlanda:2004}
can then be used to estimate the isotropic radiated energy of
E$_{\rm iso}$ of  $\sim$ 5 $\times 10^{52}$erg. 

Recently \citet{2005atteia} reported a new pseudo-redshift indicator, improving the original methods of
\citet{atteia2003}. We calculated the new pseudo-redshift using the ratio between the bolometric luminosity over the brightest 15~s and the observed E$_{\rm peak}$.  A value of X$=$1.3 was determined, which corresponds to a value of $z \sim$ 0.3.
We regard this value as a lower limit because $\alpha=-1.0$ was used for calibration and is very different from 
the value of $\alpha$ for the main emission peak of   $-$1.5. 
The method \citet{yon:2004}  yields a value of z$=$0.7$\pm$0.3 which is lower 
than than the Amati value but still consistent within the combined uncertainties.
{However there is clearly a large dispersion in pseudo-redshift
values obtained}.
Using the redshift
estimated from the Amati relation, a rest-frame lag of
0.048$^{+0.017}_{-0.015}$\,s
 was determined
between 250 and 450~s post trigger, which, from previously determined trends
 \citep{nmb00}
corresponds a peak luminosity of
$0.22^{+0.11}_{-0.07} \times\,10^{53}$\,erg s$^{-1}$. 
{\citet{barkov2006} suggest that the infrared afterglow is the
result of dust re-radiation in the envelope surrounding the GRB source and using this 
model they obtain a lower value of the redshift $\leq$0.12.}

All the indirect redshift, energy  and luminosity indicators {imply} that
GRB~041219a is a very luminous burst with isotropic output
of about $\sim$5$\times$10$^{52}$ erg. 
It should be noted that the value of $\alpha$, E$_{\rm peak}$ and the lag vary throughout this well observed burst, particularly between the precursor and the main emission phase.

\subsection{Broadband spectra}
The broadband spectra of the precursor pulse and the weak pulse at $\sim$60 s in the long quiescent interval are given in Fig. 4.  The ASM observations (Fig. 2) start at the end of the precursor pulse.  The BAT and SPI data show that the spectrum of the precursor peaked at about 2 $\times 10^{19}$ Hz 
($\sim$83~keV) and is quite different from the broadband spectra in the two subsequent intervals from 8 to 65~s and 66 to 120~s where there is strong emission in the x-ray band.  The dominance of the emission in the x-ray band is apparent in the ASM and BAT profiles in Fig. 2.

In contrast the broadband spectra over the four intervals (A to D in Fig 6) are remarkably similar because the simultaneous optical emission is correlated with the gamma ray burst.  This correlation would arise naturally if the emission in both bands were generated by a common mechanism.  The most likely mechanism is that the optical emission is the low energy tail of the synchrotron radiation generated by internal shocks in the outflow 
\citep[e.g.][]{1994katz,1999mes}.
{\citet{Fan2005}  and \citet{Fan2004} 
have suggested that the optical/near IR flash 
could be the result of emission from neutron-rich internal shocks 
colliding at a larger distance from the central engine. }
 There are simultaneous J, H and K infrared observations in interval D 
 {that are an extrapolation of the optical data}.  The J, H and K-band observations 12~hours after the burst (E in Fig. 6) { are similar but less intense
 and mark the transition from the burst to the afterglow}. 
 { This similarity between} the two measurements indicate that the infrared emission is due to the afterglow and it was already well developed in interval D 
and has been well modelled as the superposition of a reverse shock 
and a forward shock component \citep{Fan2005}.

There is an interesting contrast between the broadband spectra of GRB~041219a and GRB~050904 (F in Fig. 6) because of the large difference 
between the simultaneous optical and $\gamma$-ray emission in the two bursts.
The five GRBs that have been detected
as strong optical sources
 during the prompt $\gamma$-ray emission are GRB~990123 at $z$ = 1.60 \citep{akerlof99}, GRB~041219a with no spectroscopic redshift, GRB~050401 at $z =$ 2.9 \citep{Rykoff2005}, GRB~050904 at $z =$ 6.29 \citep{boer2006,2006watson,tag2005,cum2005, wei2005}
 and GRB~0601111b with no spectroscopic redshift \citep{klotz2006}.  The flux densities at the maximum of the optical emission are respectively 1080 mJy (V-band), 3.1 mJy (R-band), 2.6 mJy (R-band), 1300 mJy (V-band)
and $\sim$360~mJy (clear filter) if it is assumed that all GRBs are at the same redshift of 1.6 for GRB 990123 \citep{boer2006}.  At $z =$ 1.6 the optical emission from GRB~990123, GRB~050904
and GRB~0601111b are comparable and exceed that from GRB~041219a and GRB~050401 by a factor of several hundred.
The bright optical emission from GRB~990123, GRB~050904
and GRB~0601111b has been attributed to the reverse shock.  This powerful emission is not dominant in GRB~041219a and GRB~050401 because it would have overwhelmed the observed emission.  Furthermore the strong optical emission from the reverse shock is not a common feature of most GRBs \citep{Roming2005}.  
In the additional case of GRB~060124 at a redshift of  2.297 \citep{nestor2006,Cenko2006,Prochaska2006} recent simultaneous optical and
$\gamma$-ray observations with  \textit{Swift} reveal prompt optical emission with a peak of 16.88 magnitude in the V-Band (0.55mJy)
\citep{Holland2006}. The  UVOT instrument was in image mode at early times and high time resolution data was not available.  
   \citet{ramano2005} rule out a reverse shock based on the level of the optical emission and furthermore, since the optical data do not clearly track the $\gamma$-ray emission, they  suggest that external shocks may be the source of the optical emission.
The long duration and  nearby (z=0.0331 \citep{nestor060218}) weak burst/x-ray flash GRB~060218/SN 2006aj has simultaneous optical and $\gamma$-ray observations with \textit{Swift} \citep{cusumano060218,campana2006}. The peak of the prompt optical emission is only 0.118 mJy in the V band 
 \citep{campana2006}
 and is smaller than the bright optical emission from GRB~990123 by a factor of $\sim10^{8}$ (if both bursts were at a redshift of z=1.6) yielding a much larger ratio in prompt optical than 
 in the prompt $\gamma$-ray emission.

These results underline the need for continued broadband observations of the prompt emission phase and the early afterglow.

\section{Conclusions}
GRB~041219a is the brightest burst localised by \textit{INTEGRAL}
and has a
peak flux (20\,keV--8\,MeV, 1\,s integration) of
43\,ph\,cm$^{-2}$\,s$^{-1}$ ( 1.84$\times 10^{-5}$ ergs\,cm$^{-2}$\,s$^{-1}$),
which places it among the brightest bursts.
Furthermore, the T$_{\rm 90}$ duration of GRB~041219a
is 186\,s ($\sim$20\,keV--8\,MeV)  making it longer
than all but a small number of BATSE bursts. 
The main burst occurred
after a long quiescent interval of $\sim$250~s which enabled optical and near infrared telescopes to observe the burst while the
prompt event was still in progress.

We have presented comprehensive results on the temporal and spectral analyses
of GRB~041219a
(Fig.~\ref{spilc}, Fig.~\ref{bbpo_fig}), including line and afterglow searches  using 
the high resolution Germanium  spectrometer, SPI, aboard \textit{INTEGRAL}.
Spectra for the burst and sub--intervals were fit by the Band model and also by the quasithermal model
(Table~\ref{spec} and Table~\ref{bbpo_spec}). The high resolution
Germanium spectrometer data were searched for emission and absorption features and  for $\gamma$-ray afterglow. 
No significant  emission or absorption features were found and limits  of 900~eV and 120~eV are set on the most significant features.
No $\gamma$-ray afterglow was detected from the end of the prompt phase up to $\sim$12~hours post-burst (Fig.~\ref{limitss}). 
The spectral lag was determined using data from the BAT 
and changes throughout the burst (Table~\ref{spec}).

We availed of public data from BAT on \textit{Swift} and obtained 
ASM observations of \textit{Rossi X-ray Timing Explorer} 
(Fig.~\ref{BATcurve}), optical and
near-infrared observations published by \citet{vestrand:2005}
and \citet{blake:2005}. 
Broadband spectra  during 7 intervals in the prompt phase are
presented (Fig.~\ref{asmSED} and Fig.~\ref{SED})
and  compared to the high-redshift GRB~050904. 
The $\gamma$-ray and optical data in GRB~041219a 
are correlated in contrast with the large difference between 
the simultaneous optical and $\gamma$-ray emission 
in GRB~050904.

The results presented here highlight the need for continued
broadband observations of $\gamma$-ray burst and the
afterglow.

\section{Acknowledgments}
SMB acknowledges the support of the European Union through a Marie Curie Intra-European Fellowship within the Sixth Framework Program.

\hyphenation{Post-Script Sprin-ger}


\begin{thebibliography}{100}
\expandafter\ifx\csname natexlab\endcsname\relax\def\natexlab#1{#1}\fi

\bibitem[{{Akerlof} {et~al.}(1999){Akerlof}, {Balsano}, {Barthelemy}, {Bloch},
  {Butterworth}, {Casperson}, {Cline}, {Fletcher}, {Frontera}, {Gisler},
  {Heise}, {Hills}, {Kehoe}, {Lee}, {Marshall}, {McKay}, {Miller}, {Piro},
  {Priedhorsky}, {Szymanski}, \& {Wren}}]{akerlof99}
{Akerlof}, C., {Balsano}, R., {Barthelemy}, S., {et~al.} 1999, Nature, 398, 400

\bibitem[{{Amati} {et~al.}(2002){Amati}, {Frontera}, {Tavani},
  {et~al.}}]{aft+02}
{Amati}, L., {Frontera}, F., {Tavani}, M., {et~al.} 2002, A\&A, 390, 81

\bibitem[{{Atteia}(2003)}]{atteia2003}
{Atteia}, J.-L. 2003, A\&A, 407, L1

\bibitem[{{Band} {et~al.}(1993){Band}, {Matteson}, {Ford}, {Schaefer},
  {Palmer}, {Teegarden}, {Cline}, {Briggs}, {Paciesas}, {Pendleton}, {Fishman},
  {Kouveliotou}, {Meegan}, {Wilson}, \& {Lestrade}}]{band:1993}
{Band}, D., {Matteson}, J., {Ford}, L., {et~al.} 1993, ApJ, 413, 281

\bibitem[{{Band} {et~al.}(1995){Band}, {Ford}, {Matteson}, {Briggs},
  {Paciesas}, {Pendleton}, {Preece}, {Palmer}, {Teegarden}, \&
  {Schaefer}}]{bandetal95}
{Band}, D.~L., {Ford}, L.~A., {Matteson}, J.~L., {et~al.} 1995, ApJ, 447, 289

\bibitem[{{Barkov} \& {Bisnovatyi-Kogan}(2005)}]{barkov2006}
{Barkov}, M.~V. \& {Bisnovatyi-Kogan}, G.~S. 2005, [astro-ph/0503414]

\bibitem[{Barthelmy {et~al.}(2004)Barthelmy, Burrows, Cummings,
  {et~al.}}]{bbc+2004}
Barthelmy, S., Burrows, D., Cummings, J., {et~al.} 2004, {GCN 2874}

\bibitem[{{Blake} {et~al.}(2005){Blake}, {Bloom}, {Starr}, {Falco},
  {Skrutskie}, {Fenimore}, {Duch{\^ e}ne}, {Szentgyorgyi}, {Hornstein},
  {Prochaska}, {McCabe}, {Ghez}, {Konopacky}, {Stapelfeldt}, {Hurley},
  {Campbell}, {Kassis}, {Chaffee}, {Gehrels}, {Barthelmy}, {Cummings},
  {Hullinger}, {Krimm}, {Markwardt}, {Palmer}, {Parsons}, {McLean}, \&
  {Tueller}}]{blake:2005}
{Blake}, C.~H., {Bloom}, J.~S., {Starr}, D.~L., {et~al.} 2005, Nature, 435, 181

\bibitem[{{Bo{\"e}r} {et~al.}(2006){Bo{\"e}r}, {Atteia}, {Damerdji}, {Gendre},
  {Klotz}, \& {Stratta}}]{boer2006}
{Bo{\"e}r}, M., {Atteia}, J.~L., {Damerdji}, Y., {et~al.} 2006, ApJL, 638, L71

\bibitem[{{Bosnjak} {et~al.}(2006){Bosnjak}, {Celotti}, \&
  {Ghirlanda}}]{bosnjak2006}
{Bosnjak}, Z., {Celotti}, A., \& {Ghirlanda}, G. 2006, [astro-ph/0604425]

\bibitem[{{B{\"o}ttcher}(2004)}]{bott04}
{B{\"o}ttcher}, M. 2004, Advances in Space Research, 34, 2696

\bibitem[{{Briggs} {et~al.}(1998){Briggs}, {Pendleton}, {Brainerd},
  {Connaughton}, {Kippen}, {Meegan}, \& {Hurley}}]{briggsetal97}
{Briggs}, M.~S., {Pendleton}, G.~N., {Brainerd}, J.~J., {et~al.} 1998, in
  American Institute of Physics Conference Series, ed. C.~A. {Meegan}, R.~D.
  {Preece}, \& T.~M. {Koshut}, 104

\bibitem[{{Campana} {et~al.}(2006){Campana}, {Mangano}, {Blustin}, {Brown}, \&
  {Burrows}}]{campana2006}
{Campana}, S., {Mangano}, V., {Blustin}, A.~J., {Brown}, P., \& {Burrows},
  D.~N. 2006, [astro-ph/0603279 ]

\bibitem[{{Cenko} {et~al.}(2006){Cenko}, {Berger}, \& {Cohen}}]{Cenko2006}
{Cenko}, S.~B., {Berger}, E., \& {Cohen}, J. 2006, GRB Coordinates Network,
  4592, 1

\bibitem[{{Chen} {et~al.}(2005){Chen}, {Lou}, {Wu}, {Qu}, {Jia}, \&
  {Yang}}]{clw+05}
{Chen}, L., {Lou}, Y.-Q., {Wu}, M., {et~al.} 2005, ApJ, 619, 983

\bibitem[{{Costa} {et~al.}(1997){Costa}, {Frontera}, {Heise}, {Feroci}, {in 't
  Zand}, {Fiore}, {Cinti}, {dal Fiume}, {Nicastro}, {Orlandini}, {Palazzi},
  {Rapisarda}, {Zavattini}, {Jager}, {Parmar}, {Owens}, {Molendi}, {Cusumano},
  {Maccarone}, {Giarrusso}, {Coletta}, {Antonelli}, {Giommi}, {Muller}, {Piro},
  \& {Butler}}]{costa:1997}
{Costa}, E., {Frontera}, F., {Heise}, J., {et~al.} 1997, Nature, 387, 783

\bibitem[{{Crider} {et~al.}(1997){Crider}, {Liang}, {Smith}, {Preece},
  {Briggs}, {Pendleton}, {Paciesas}, {Band}, \& {Matteson}}]{crider1997}
{Crider}, A., {Liang}, E.~P., {Smith}, I.~A., {et~al.} 1997, ApJL, 479, L39

\bibitem[{{Cusumano} {et~al.}(2006){Cusumano}, {Barthelmy}, {Gehrels},
  {Hunsberger}, {Immler}, {Marshall}, {Palmer}, \& {Sakamoto}}]{cusumano060218}
{Cusumano}, G., {Barthelmy}, S., {Gehrels}, N., {et~al.} 2006, GRB Coordinates
  Network, 4775, 1

\bibitem[{{Cusumano} {et~al.}(2005){Cusumano}, {Mangano}, {Chincarini},
  {Panaitescu}, {Burrows}, {et~al.}}]{cum2005}
{Cusumano}, G., {Mangano}, V., {Chincarini}, G., {et~al.} 2005,
  [astro-ph/0509737]

\bibitem[{{Diehl} {et~al.}(2003){Diehl}, {Baby}, {Beckmann}, {Connell},
  {Dubath}, {Jean}, {Kn{\"o}dlseder}, {Roques}, {Schanne}, {Shrader},
  {Skinner}, {Strong}, {Sturner}, {Teegarden}, {von Kienlin}, \&
  {Weidenspointner}}]{Diehl2003}
{Diehl}, R., {Baby}, N., {Beckmann}, V., {et~al.} 2003, A\&A, 411, L117

\bibitem[{{Drago} \& {Pagliara}(2005)}]{2005Drago}
{Drago}, A. \& {Pagliara}, G. 2005, [astro-ph/0512602]

\bibitem[{{Fan} \& {Wei}(2004)}]{Fan2004}
{Fan}, Y.~Z. \& {Wei}, D.~M. 2004, ApJL, 615, L69

\bibitem[{{Fan} {et~al.}(2005){Fan}, {Zhang}, \& {Wei}}]{Fan2005}
{Fan}, Y.~Z., {Zhang}, B., \& {Wei}, D.~M. 2005, ApJL, 628, L25

\bibitem[{Fenimore {et~al.}(2004)Fenimore, Barthelmy, Cummings,
  {et~al.}}]{fbc+2004}
Fenimore, E., Barthelmy, S., Cummings, J., {et~al.} 2004, {GCN 2906}

\bibitem[{Fishman {et~al.}(1994)Fishman, Meegan, Wilson, {et~al.}}]{fish1994}
Fishman, G.~J., Meegan, C.~A., Wilson, R.~B., {et~al.} 1994, ApJS, 92, 229

\bibitem[{{Gehrels} {et~al.}(2004){Gehrels}, {Chincarini}, {Giommi}, {Mason},
  {Nousek}, {Wells}, {White}, {Barthelmy}, {Burrows}, {Cominsky}, {Hurley},
  {Marshall}, {M{\' e}sz{\' a}ros}, {Roming}, {Angelini}, {Barbier}, {Belloni},
  {Campana}, {Caraveo}, {Chester}, {Citterio}, {Cline}, {Cropper}, {Cummings},
  {Dean}, {Feigelson}, {Fenimore}, {Frail}, {Fruchter}, {Garmire}, {Gendreau},
  {Ghisellini}, {Greiner}, {Hill}, {Hunsberger}, {Krimm}, {Kulkarni}, {Kumar},
  {Lebrun}, {Lloyd-Ronning}, {Markwardt}, {Mattson}, {Mushotzky}, {Norris},
  {Osborne}, {Paczynski}, {Palmer}, {Park}, {Parsons}, {Paul}, {Rees},
  {Reynolds}, {Rhoads}, {Sasseen}, {Schaefer}, {Short}, {Smale}, {Smith},
  {Stella}, {Tagliaferri}, {Takahashi}, {Tashiro}, {Townsley}, {Tueller},
  {Turner}, {Vietri}, {Voges}, {Ward}, {Willingale}, {Zerbi}, \&
  {Zhang}}]{gehrels:2004}
{Gehrels}, N., {Chincarini}, G., {Giommi}, P., {et~al.} 2004, ApJ, 611, 1005

\bibitem[{{Ghirlanda} {et~al.}(2002){Ghirlanda}, {Celotti}, \&
  {Ghisellini}}]{ghirlanda2002}
{Ghirlanda}, G., {Celotti}, A., \& {Ghisellini}, G. 2002, A\&A, 393, 409

\bibitem[{{Ghirlanda} {et~al.}(2003){Ghirlanda}, {Celotti}, \&
  {Ghisellini}}]{ghirlanda2003}
{Ghirlanda}, G., {Celotti}, A., \& {Ghisellini}, G. 2003, A\&A, 406, 879

\bibitem[{{Ghirlanda} {et~al.}(2004){Ghirlanda}, {Ghisellini}, \&
  {Lazzati}}]{ghirlanda:2004}
{Ghirlanda}, G., {Ghisellini}, G., \& {Lazzati}, D. 2004, ApJ, 616, 331

\bibitem[{{Golenetskii} {et~al.}(2006){Golenetskii}, {Aptekar}, {Mazets},
  {Pal'Shin}, {Frederiks}, \& {Cline}}]{Golenstskii2006}
{Golenetskii}, S., {Aptekar}, R., {Mazets}, E., {et~al.} 2006, GRB Coordinates
  Network, 4599, 1

\bibitem[{{G\"{o}tz} {et~al.}(2004){G\"{o}tz}, Mereghetti, Shaw,
  {et~al.}}]{gms+2004}
{G\"{o}tz}, D., Mereghetti, S., Shaw, S., {et~al.} 2004, {GCN 2866}

\bibitem[{{Hakkila} \& {Giblin}(2004)}]{hakkila2004}
{Hakkila}, J. \& {Giblin}, T.~W. 2004, ApJ, 610, 361

\bibitem[{{Hjorth} {et~al.}(2003){Hjorth}, {Sollerman}, {M{\o}ller}, {Fynbo},
  {Woosley}, {Kouveliotou}, {Tanvir}, {Greiner}, {Andersen}, {Castro-Tirado},
  {Castro Cer{\' o}n}, {Fruchter}, {Gorosabel}, {Jakobsson}, {Kaper}, {Klose},
  {Masetti}, {Pedersen}, {Pedersen}, {Pian}, {Palazzi}, {Rhoads}, {Rol}, {van
  den Heuvel}, {Vreeswijk}, {Watson}, \& {Wijers}}]{hjorth2003}
{Hjorth}, J., {Sollerman}, J., {M{\o}ller}, P., {et~al.} 2003, Nature, 423, 847

\bibitem[{{Holland} {et~al.}(2006{\natexlab{a}}){Holland}, {Barthelmy},
  {Burrows}, {Gehrels}, {Hunsberger}, {Kennea}, {La Parola}, {Markwardt},
  {Page}, {Palmer}, \& {Sakamoto}}]{Holland060124}
{Holland}, S.~T., {Barthelmy}, S., {Burrows}, D.~N., {et~al.}
  2006{\natexlab{a}}, GRB Coordinates Network, 4570, 1

\bibitem[{{Holland} {et~al.}(2006{\natexlab{b}}){Holland}, {Smith}, {Huckle},
  \& {Gehrels}}]{Holland2006}
{Holland}, S.~T., {Smith}, P., {Huckle}, H., \& {Gehrels}, N.
  2006{\natexlab{b}}, GRB Coordinates Network, 4580, 1

\bibitem[{{Kaneko} {et~al.}(2003){Kaneko}, {Preece}, \& {Briggs}}]{Kaneko2003}
{Kaneko}, Y., {Preece}, R.~D., \& {Briggs}, M.~S. 2003, American Astronomical
  Society Meeting Abstracts, 203

\bibitem[{{Katz}(1994)}]{1994katz}
{Katz}, J.~I. 1994, ApJL, 432, L107

\bibitem[{{Klotz} {et~al.}(2006){Klotz}, {Gentre}, {Stratta}, {Atteia},
  {Bo{\"e}r}, {Malacrino}, {Damerdji}, \& {Behrend}}]{klotz2006}
{Klotz}, A., {Gentre}, B., {Stratta}, G., {et~al.} 2006, [astro-ph/0604061]

\bibitem[{{Kocevski} \& {Liang}(2003)}]{kl03}
{Kocevski}, D. \& {Liang}, E. 2003, ApJ, 594, 385

\bibitem[{{Kocevski} {et~al.}(2003){Kocevski}, {Ryde}, \& {Liang}}]{dan2003}
{Kocevski}, D., {Ryde}, F., \& {Liang}, E. 2003, ApJ, 596, 389

\bibitem[{{Koshut} {et~al.}(1995){Koshut}, {Kouveliotou}, {Paciesas}, {van
  Paradijs}, {Pendleton}, {Briggs}, {Fishman}, \& {Meegan}}]{koshut:1995}
{Koshut}, T.~M., {Kouveliotou}, C., {Paciesas}, W.~S., {et~al.} 1995, ApJ,
  452, 145

\bibitem[{{Lazzati}(2005)}]{2005Laz}
{Lazzati}, D. 2005, MNRAS, 357, 722

\bibitem[{{Lazzati} {et~al.}(2000){Lazzati}, {Ghisellini}, {Celotti}, \&
  {Rees}}]{2000lazzati}
{Lazzati}, D., {Ghisellini}, G., {Celotti}, A., \& {Rees}, M.~J. 2000, ApJL,
  529, L17

\bibitem[{{Levine} \& {Remillard}(2004)}]{ASMGCN}
{Levine}, A. \& {Remillard}, R. 2004, GRB Coordinates Network, 2917, 1

\bibitem[{{Levine} {et~al.}(1996){Levine}, {Bradt}, {Cui}, {Jernigan},
  {Morgan}, {Remillard}, {Shirey}, \& {Smith}}]{LEvine1996}
{Levine}, A.~M., {Bradt}, H., {Cui}, W., {et~al.} 1996, ApJL, 469, L33

\bibitem[{{Liang}(1997)}]{1997liang}
{Liang}, E.~P. 1997, ApJL, 491, L15

\bibitem[{{Lloyd} \& {Petrosian}(2000)}]{lloyd2000}
{Lloyd}, N.~M. \& {Petrosian}, V. 2000, ApJ, 543, 722

\bibitem[{{Lloyd-Ronning} \& {Ramirez-Ruiz}(2002)}]{LL_RR:2002}
{Lloyd-Ronning}, N.~M. \& {Ramirez-Ruiz}, E. 2002, ApJ, 576, 101

\bibitem[{{Mazets} {et~al.}(1980){Mazets}, {Golenetskii}, {Aptekar}, {Guryan},
  \& {Ilinskii}}]{maz80}
{Mazets}, E.~P., {Golenetskii}, S.~V., {Aptekar}, R.~L., {Guryan}, Y.~A., \&
  {Ilinskii}, V.~N. 1980, Soviet Astronomy Letters, 6, 372

\bibitem[{{McBreen} {et~al.}(2002{\natexlab{a}}){McBreen}, {McBreen}, {Hanlon},
  \& {Quilligan}}]{mmhq2002}
{McBreen}, S., {McBreen}, B., {Hanlon}, L., \& {Quilligan}, F.
  2002{\natexlab{a}}, A\&A, 393, L29

\bibitem[{{McBreen} {et~al.}(2002{\natexlab{b}}){McBreen}, {McBreen},
  {Quilligan}, \& {Hanlon}}]{mmqh02}
{McBreen}, S., {McBreen}, B., {Quilligan}, F., \& {Hanlon}, L.
  2002{\natexlab{b}}, A\&A, 385, L19

\bibitem[{{McGlynn} {et~al.}(2005){McGlynn}, {McBreen}, {Hanlon}, {McBreen},
  {Moran}, \& {Von Kienlin}}]{sinead2005}
{McGlynn}, S., {McBreen}, S., {Hanlon}, L., {et~al.} 2005, [astro-ph/0505349]

\bibitem[{{Medvedev}(2000)}]{med2000}
{Medvedev}, M.~V. 2000, ApJ, 540, 704

\bibitem[{Mereghetti {et~al.}(2003)Mereghetti, {G\"{o}tz}, Borkowski, Walter,
  \& Pedersen}]{mgbwp03}
Mereghetti, S., {G\"{o}tz}, D., Borkowski, J., Walter, R., \& Pedersen, H.
  2003, A\&A, 411, L291

\bibitem[{{M{\'e}sz{\'a}ros} \& {Rees}(1999)}]{1999mes}
{M{\'e}sz{\'a}ros}, P. \& {Rees}, M.~J. 1999, MNRAS, 306, L39

\bibitem[{{M{\'e}sz{\'a}ros} \& {Rees}(2000)}]{mes2000}
{M{\'e}sz{\'a}ros}, P. \& {Rees}, M.~J. 2000, ApJ, 530, 292

\bibitem[{{Mirabal} \& {Halpern}(2006{\natexlab{a}})}]{nestor2006}
{Mirabal}, N. \& {Halpern}, J.~P. 2006{\natexlab{a}}, GRB Coordinates Network,
  4591, 1

\bibitem[{{Mirabal} \& {Halpern}(2006{\natexlab{b}})}]{nestor060218}
{Mirabal}, N. \& {Halpern}, J.~P. 2006{\natexlab{b}}, GRB Coordinates Network,
  4792, 1

\bibitem[{{Montroll} \& {Shlesinger}(1982)}]{montshles:1982}
{Montroll}, E.~W. \& {Shlesinger}, M.~F. 1982, Proceedings of the National
  Academy of Sciences of the United States of America-Physical Sciences, 79,
  3380

\bibitem[{{Moran} {et~al.}(2005){Moran}, {Mereghetti}, {G\"otz}, {Hanlon}, {von
  Kienlin}, {McBreen}, {et~al.}}]{moran2004}
{Moran}, L., {Mereghetti}, S., {G\"otz}, D., {et~al.} 2005, A\&A, 432, 467

\bibitem[{{Murakami} {et~al.}(1988){Murakami}, {Fujii}, {Hayashida}, {Itoh}, \&
  {Nishimura}}]{murakamietal88}
{Murakami}, T., {Fujii}, M., {Hayashida}, K., {Itoh}, M., \& {Nishimura}, J.
  1988, Nature, 335, 234

\bibitem[{{Murakami} {et~al.}(1991){Murakami}, {Inoue}, {Nishimura}, {van
  Paradijs}, \& {Fenimore}}]{murk:1991}
{Murakami}, T., {Inoue}, H., {Nishimura}, J., {van Paradijs}, J., \&
  {Fenimore}, E.~E. 1991, Nature, 350, 592

\bibitem[{{Nakar} \& {Piran}(2002)}]{nakar:2002}
{Nakar}, E. \& {Piran}, T. 2002, ApJL, 572, L139

\bibitem[{{Norris} {et~al.}(2000){Norris}, {Marani}, \& {Bonnell}}]{nmb00}
{Norris}, J.~P., {Marani}, G.~F., \& {Bonnell}, J.~T. 2000, ApJ, 534, 248

\bibitem[{{Paciesas} {et~al.}(1999){Paciesas}, {Meegan}, {Pendleton}, {Briggs},
  {Kouveliotou}, {Koshut}, {Lestrade}, {McCollough}, {Brainerd}, {Hakkila},
  {Henze}, {Preece}, {Connaughton}, {Kippen}, {Mallozzi}, {Fishman},
  {Richardson}, \& {Sahi}}]{Paciesas:1999}
{Paciesas}, W.~S., {Meegan}, C.~A., {Pendleton}, G.~N., {et~al.} 1999, ApJS,
  122, 465

\bibitem[{{Page} {et~al.}(2005){Page}, {Burrows}, {Beardmore}, {Palmer},
  {Kennea}, {Gehrels}, {Markwardt}, {Page}, {Sakamoto}, {Chester}, \&
  {Boyd}}]{2005Page}
{Page}, M., {Burrows}, D., {Beardmore}, A., {et~al.} 2005, GRB Coordinates
  Network, 3830, 1

\bibitem[{{Pal'Shin} \& {Frederiks}(2005)}]{golenetskii2005}
{Pal'Shin}, V. \& {Frederiks}, D. 2005, GRB Coordinates Network, 3852, 1

\bibitem[{{P\/elangeon} {et~al.}(2005){P\/elangeon}, {Atteia}, {Lamb}, \&
  {Ricker}}]{2005atteia}
{P\/elangeon}, A., {Atteia}, J.~L., {Lamb}, D.~Q., \& {Ricker}, G.~R. 2005,
  [astro-ph/0601150]

\bibitem[{Piran(2004)}]{piran:1143}
Piran, T. 2004, Reviews of Modern Physics, 76, 1143

\bibitem[{{Preece} {et~al.}(2002){Preece}, {Briggs}, {Giblin}, {Mallozzi},
  {Pendleton}, {Paciesas}, \& {Band}}]{preece_sync}
{Preece}, R.~D., {Briggs}, M.~S., {Giblin}, T.~W., {et~al.} 2002, ApJ, 581,
  1248

\bibitem[{Preece {et~al.}(2000)Preece, Briggs, Mallozzi, {et~al.}}]{pbm+00}
Preece, R.~D., Briggs, M.~S., Mallozzi, R., {et~al.} 2000, ApJS, 126, 19

\bibitem[{{Prochaska} {et~al.}(2006){Prochaska}, {Foley}, {Tran}, {Bloom}, \&
  {Chen}}]{Prochaska2006}
{Prochaska}, J.~X., {Foley}, R., {Tran}, H., {Bloom}, J.~S., \& {Chen}, H.-W.
  2006, GRB Coordinates Network, 4593, 1

\bibitem[{{Protassov} {et~al.}(2002){Protassov}, {van Dyk}, {Connors},
  {Kashyap}, \& {Siemiginowska}}]{prot2002}
{Protassov}, R., {van Dyk}, D.~A., {Connors}, A., {Kashyap}, V.~L., \&
  {Siemiginowska}, A. 2002, ApJ, 571, 545

\bibitem[{{Quilligan} {et~al.}(2002){Quilligan}, {McBreen}, {Hanlon},
  {McBreen}, {Hurley}, \& {Watson}}]{quilligan:2002}
{Quilligan}, F., {McBreen}, B., {Hanlon}, L., {et~al.} 2002, A\&A, 385, 377

\bibitem[{{Ramirez-Ruiz} {et~al.}(2002){Ramirez-Ruiz}, {MacFadyen}, \&
  {Lazzati}}]{rr_macf_lazz:2002}
{Ramirez-Ruiz}, E., {MacFadyen}, A.~I., \& {Lazzati}, D. 2002, MNRAS, 331, 197

\bibitem[{{Ramirez-Ruiz} \& {Merloni}(2001)}]{RR_1:2000}
{Ramirez-Ruiz}, E. \& {Merloni}, A. 2001, MNRAS, 320, L25

\bibitem[{{Ramirez-Ruiz} {et~al.}(2001){Ramirez-Ruiz}, {Merloni}, \&
  {Rees}}]{RR_2:2000}
{Ramirez-Ruiz}, E., {Merloni}, A., \& {Rees}, M.~J. 2001, MNRAS, 324, 1147

\bibitem[{{Rees} \& {M{\'e}sz{\'a}ros}(2005)}]{rees2005}
{Rees}, M.~J. \& {M{\'e}sz{\'a}ros}, P. 2005, ApJ, 628, 847

\bibitem[{{Romano} {et~al.}(2006){Romano}, {Campana}, {Chincarini}, {Cummings},
  {Cusumano}, \& others.}]{ramano2005}
{Romano}, P., {Campana}, S., {Chincarini}, G., {et~al.} 2006,
  [astro-ph/0602497]

\bibitem[{{Roming} {et~al.}(2005){Roming}, {Schady}, {Fox}, {Zhang}, {Liang},
  {et~al.}}]{Roming2005}
{Roming}, P.~W.~A., {Schady}, P., {Fox}, D.~B., {et~al.} 2005,
  [astro-ph/0509273]

\bibitem[{{Ryde}(2004)}]{ryde2004}
{Ryde}, F. 2004, ApJ, 614, 827

\bibitem[{{Ryde}(2005{\natexlab{a}})}]{ryde05}
{Ryde}, F. 2005{\natexlab{a}}, A\&A, 429, 869

\bibitem[{{Ryde}(2005{\natexlab{b}})}]{ryde05b}
{Ryde}, F. 2005{\natexlab{b}}, ApJL, 625, L95

\bibitem[{{Ryde} {et~al.}(2005){Ryde}, {Kocevski}, {Bagoly}, {Ryde}, \&
  {M{\'e}sz{\'a}ros}}]{Ryde_Kov:2005}
{Ryde}, F., {Kocevski}, D., {Bagoly}, Z., {Ryde}, N., \& {M{\'e}sz{\'a}ros}, A.
  2005, A\&A, 432, 105

\bibitem[{{Rykoff} {et~al.}(2005){Rykoff}, {Yost}, {Krimm}, {Aharonian},
  {Akerlof}, {Alatalo}, {Ashley}, {Barthelmy}, {Gehrels}, {G{\"o}{\v g}{\"u}{\c
  s}}, {G{\"u}ver}, {Horns}, {K{\i}z{\i}lo{\v g}lu}, {McKay}, {{\"O}zel},
  {Phillips}, {Quimby}, {Rujopakarn}, {Schaefer}, {Smith}, {Swan}, {Vestrand},
  {Wheeler}, \& {Wren}}]{Rykoff2005}
{Rykoff}, E.~S., {Yost}, S.~A., {Krimm}, H.~A., {et~al.} 2005, ApJL, 631, L121

\bibitem[{{Sazonov} {et~al.}(2004){Sazonov}, {Lutovinov}, \& {Sunyaev}}]{sls04}
{Sazonov}, S.~Y., {Lutovinov}, A.~A., \& {Sunyaev}, R.~A. 2004, Nature, 430, 646

\bibitem[{{Soderberg} \& {Frail}(2004)}]{Soderberg04}
{Soderberg}, A.~M. \& {Frail}, D.~A. 2004, GCN 2881

\bibitem[{{Sonoda} {et~al.}(2004){Sonoda}, {Maeno}, {Matsuo}, \&
  {Yamauchi}}]{2004Sonda}
{Sonoda}, E., {Maeno}, S., {Matsuo}, Y., \& {Yamauchi}, M. 2004, GRB
  Coordinates Network, 2882

\bibitem[{{Tagliaferri} {et~al.}(2005){Tagliaferri}, {Antonelli}, {Chincarini},
  {Fern{\'a}ndez-Soto}, {Malesani}, {Della Valle}, {D'Avanzo}, {Grazian},
  {Testa}, {Campana}, {Covino}, {Fiore}, {Stella}, {Castro-Tirado},
  {Gorosabel}, {Burrows}, {Capalbi}, {Cusumano}, {Conciatore}, {D'Elia},
  {Filliatre}, {Fugazza}, {Gehrels}, {Goldoni}, {Guetta}, {Guziy}, {Held},
  {Hurley}, {Israel}, {Jel{\'{\i}}nek}, {Lazzati}, {L{\'o}pez-Echarri},
  {Melandri}, {Mirabel}, {Moles}, {Moretti}, {Mason}, {Nousek}, {Osborne},
  {Pellizza}, {Perna}, {Piranomonte}, {Piro}, {de Ugarte Postigo}, \&
  {Romano}}]{tag2005}
{Tagliaferri}, G., {Antonelli}, L.~A., {Chincarini}, G., {et~al.} 2005, A\&A,
  443, L1

\bibitem[{{van der Horst} {et~al.}(2004{\natexlab{a}}){van der Horst}, {Rol},
  \& {Strom}}]{van1:04}
{van der Horst}, A.~J., {Rol}, E., \& {Strom}, R. 2004{\natexlab{a}}, GCN 2894

\bibitem[{{van der Horst} {et~al.}(2004{\natexlab{b}}){van der Horst}, {Rol},
  \& {Strom}}]{van2:04}
{van der Horst}, A.~J., {Rol}, E., \& {Strom}, R. 2004{\natexlab{b}}, GCN 2895

\bibitem[{{Vanderspek} {et~al.}(2004){Vanderspek}, {Sakamoto}, {Barraud},
  {Tamagawa}, {Graziani}, {Suzuki}, {Shirasaki}, {Prigozhin}, {Villasenor},
  {Jernigan}, {Crew}, {Atteia}, {Hurley}, {Kawai}, {Lamb}, {Ricker}, {Woosley},
  {Butler}, {Doty}, {Dullighan}, {Donaghy}, {Fenimore}, {Galassi}, {Matsuoka},
  {Takagishi}, {Torii}, {Yoshida}, {Boer}, {Dezalay}, {Olive}, {Braga},
  {Manchanda}, \& {Pizzichini}}]{2004Van}
{Vanderspek}, R., {Sakamoto}, T., {Barraud}, C., {et~al.} 2004, ApJ, 617, 1251

\bibitem[{Vedrenne {et~al.}(2003)Vedrenne, Roques, {Sch\"{o}nfelder},
  {et~al.}}]{ved2003}
Vedrenne, G., Roques, J.~P., {Sch\"{o}nfelder}, V., {et~al.} 2003, A\&A, 411,
  L63

\bibitem[{{Vestrand} {et~al.}(2005){Vestrand}, {Wozniak}, {Wren}, {Fenimore},
  {Sakamoto}, {White}, {Casperson}, {Davis}, {Evans}, {Galassi}, {McGowan},
  {Schier}, {Asa}, {Barthelmy}, {Cummings}, {Gehrels}, {Hullinger}, {Krimm},
  {Markwardt}, {McLean}, {Palmer}, {Parsons}, \& {Tueller}}]{vestrand:2005}
{Vestrand}, W.~T., {Wozniak}, P.~R., {Wren}, J.~A., {et~al.} 2005, Nature, 435,
  178

\bibitem[{{Watson} {et~al.}(2004){Watson}, {Hjorth}, {Levan}, {Jakobsson},
  {O'Brien}, {Osborne}, {Pedersen}, {Reeves}, {Tedds}, {Vaughan}, {Ward}, \&
  {Willingale}}]{watson2004}
{Watson}, D., {Hjorth}, J., {Levan}, A., {et~al.} 2004, ApJL, 605, L101

\bibitem[{{Watson} {et~al.}(2006){Watson}, {Reeves}, {Hjorth}, {Fynbo},
  {Jakobsson}, {Pedersen}, {Sollerman}, {Cer{\'o}n}, {McBreen}, \&
  {Foley}}]{2006watson}
{Watson}, D., {Reeves}, J.~N., {Hjorth}, J., {et~al.} 2006, ApJL, 637, L69

\bibitem[{{Wei} {et~al.}(2005){Wei}, {Yan}, \& {Fan}}]{wei2005}
{Wei}, D.~M., {Yan}, T., \& {Fan}, Y. 2005, [astro-ph/0511154]

\bibitem[{Winkler {et~al.}(2003)Winkler, {Courvoisier}, {Di Cocco},
  {et~al.}}]{wink2003}
Winkler, C., {Courvoisier}, T.~J.~L., {Di Cocco}, G., {et~al.} 2003, A\&A, 411,
  L1

\bibitem[{{Yonetoku} {et~al.}(2004){Yonetoku}, {Murakami}, {Nakamura},
  {Yamazaki}, {Inoue}, \& {Ioka}}]{yon:2004}
{Yonetoku}, D., {Murakami}, T., {Nakamura}, T., {et~al.} 2004, ApJ, 609, 935

\bibitem[{{Zhang} \& {M{\'e}sz{\'a}ros}(2004)}]{Zhang2004}
{Zhang}, B. \& {M{\'e}sz{\'a}ros}, P. 2004, International Journal of Modern
  Physics A, 19, 2385

\end{thebibliography}
\end{document}